\documentclass{article}
\usepackage{spconf,amsmath,graphicx,hyperref}
\usepackage{amsmath}
\usepackage{enumitem}
\usepackage{amssymb}
\usepackage{multirow}
\usepackage{url}
\title{BEST-STD 2.0: Balanced and Efficient Speech Tokenizer for Spoken Term Detection}

\name{Anup Singh$^{\S}$ \qquad Vipul Arora$^{\dagger \star}$  \qquad Kris Demuynck$^{\S \star}$  \thanks{$\star$ Equal advising.} }

\address{$^{\S}$ IDLab, Department of Electronics and Information Systems, Ghent University, Belgium \\
      $^{\dagger}$ ESAT-PSI, KU Leuven, Belgium }

\begin{document}
\ninept

\maketitle
\begin{abstract}
Fast and accurate spoken content retrieval is vital for applications such as voice search. Query-by-Example Spoken Term Detection (STD) involves retrieving matching segments from an audio database given a spoken query. Token-based STD systems, which use discrete speech representations, enable efficient search but struggle with robustness to noise and reverberation, and with inefficient token utilization. We address these challenges by proposing a noise and reverberation-augmented training strategy to improve tokenizer robustness. In addition, we introduce optimal transport-based regularization to ensure balanced token usage and enhance token efficiency. To further speed up retrieval, we adopt a TF-IDF-based search mechanism. Empirical evaluations demonstrate that the proposed method outperforms STD baselines across various distortion levels while maintaining high search efficiency.

\end{abstract}
\section{Introduction}
The rapid growth of spoken content across digital platforms has underscored the need for robust and efficient Query-by-Example Spoken Term Detection (QbE-STD) systems. However, existing systems often struggle in noisy and reverberant environments, limiting their effectiveness in real-world applications such as voice search.

The ASR-based approaches \cite{mamou2007vocabulary, miller2007rapid, wang2008comparison} represent speech using subword lattices to handle OOV terms but require highly accurate ASR models, which are difficult to train for short, low-context queries. Direct audio comparison systems \cite{ram2018cnn, ram2020neural} circumvent transcription by performing direct audio-template matching using Segmental Dynamic Time Warping (SDTW) \cite{tsai2021segmental}, but the high computational cost of SDTW hinders scalability. Other approaches \cite{he2016multi, chung2016audio, kamper2016deep} learn discriminative acoustic word embeddings, but these methods require precise word boundaries during training and inference, making them impractical for most real-world scenarios. Moreover, most existing STD systems are designed for clean acoustic environments and suffer severe degradation under noise and reverberation.

To alleviate these issues, our previous work, BEST-STD \cite{singh2025best}, introduced a discrete tokenization strategy that converts speech into subword-like units, enabling fast retrieval with text-based search algorithms, eliminating boundary annotations at inference, and efficiently handling OOV terms. However, this approach still produces tokens that are sensitive to acoustic conditions and exhibit low entropy, limiting both robustness and discriminability. Existing speech tokenizers \cite{hsu2021hubert, chen2022wavlm, zhang2023speechtokenizer} that generate semantic tokens also remain entangled with speaker characteristics and other acoustic cues.

This paper introduces \textbf{BEST-STD 2.0}, a novel speech tokenizer designed to produce speaker-agnostic and noise-robust tokens. Our framework transforms input speech into a contextual frame-level embedding sequence using a bidirectional Mamba encoder \cite{singh2025best}, followed by tokenization through vector quantization \cite{van2017neural}. To ensure consistent tokenization, we employ a self-supervised training objective that maps different utterances of the same term to the same token sequence, even under distortions like noise and reverberation. Furthermore, to address the common issue of token index collapse \cite{huh2023straightening} in discrete representation learning, we reformulate token learning as a balanced clustering problem and introduce an optimal-transport regularization \cite{cuturi2013sinkhorn} that promotes uniform codebook usage and enhances discriminability. Empirical evaluations on the LibriSpeech and TIMIT datasets with added real-world noise scenarios show that our system outperforms existing STD baselines across a range of acoustic conditions, demonstrating improved robustness and search efficiency. By enabling text-like search capabilities over raw speech, our method provides a scalable solution for spoken content retrieval. Overall, the main contributions of our work are as follows:
\begin{itemize}[leftmargin=*]
    \item A noise-augmented training framework that produces noise-robust, speaker-agnostic speech tokens.
    \item A novel application of optimal transport to prevent codebook collapse and ensure balanced token utilization.
    \item A dedicated noise-robustness loss to further strengthen token stability under adverse conditions.
\end{itemize}
\label{sec:intro}

\section{Proposed Approach }
\begin{figure}
    % \centering
    \includegraphics[width=\linewidth]{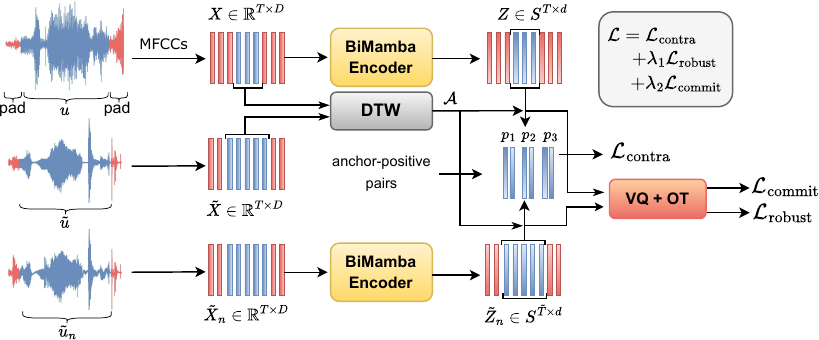}
    \caption{Illustration of our self-supervised learning framework for robust speech tokenization.}
    \label{fig:overview}
\end{figure}

The overall architecture of the proposed approach is illustrated in Fig. \ref{fig:overview}.

\subsection{Model Architecture}

We adopt the design choices for the Mamba feature encoder as described in \cite{singh2025best}. The encoder $f: X \rightarrow Z$ comprises $L$ layers of bidirectional Mamba blocks, which map the input audio feature sequence $X = \{x_1, \dots, x_t\}$ to the corresponding contextual speech representations $Z = \{z_1, \dots, z_t\}$. The output sequence embeddings of the final encoder layer are projected into a $d$-dimensional space, followed by $L_2$ normalization to ensure a unit norm.

\subsection{Self-Supervised Learning Framework}
\textbf{Robust frame-level embeddings.} For a given spoken term \( w \), we consider a pair of utterances \( (u, \tilde{u}) \) with durations \( t \) and \( \tilde{t} \) (WLOG, \( \tilde{t} > t \)), spoken by different speakers. To simulate real-world scenarios, we add contextual padding to both utterances, ensuring they serve as fixed-length inputs  to the encoder. Moreover, we stochastically add noise and reverberation to $\tilde{u}$ to generate its corresponding distorted version $\tilde{u}_n$. These padded utterances are then processed to extract their corresponding MFCC feature sequences \( X \), \( \tilde{X} \), and \( \tilde{X}_n \). 

We employ DTW to obtain the alignment $\mathcal{A}$ between \( X \) and \( \tilde{X} \). Note that during the alignment process, frames in \( X \) and \( \tilde{X} \) corresponding to the padding are excluded, yielding sequences of lengths \( T \) and \( \tilde{T} \) for alignment:
\begin{equation}
    \mathcal{A} = \{(t, \tilde{t}) \mid \tilde{t}=\arg \min_{t' \in S_t} d(X_t,\tilde{X}_{t'}), t \in [1, T], S_t \subseteq [1, \tilde{T}]\},
    \label{alignment}
\end{equation}
where $t$ denotes the frame index in $X$ and $S_t$ the set of indices in $\tilde{X}$ aligned with $X_t$, and $d(.)$ is the Euclidean distance function. Subsequently, \( X \) and \( \tilde{X}_n \) are fed into the encoder to generate their respective embedding sequences \( Z \) and \( \tilde{Z}_n \). 

The DTW-based alignment $\mathcal{A}$ serves as self-supervision, enabling the creation of frame-level anchor-positive pairs: 
\begin{equation}
    p_t = (z_t, \tilde{z}_{n_{\tilde{t}}}), \textrm{where}\, (t, \tilde{t}) \in \mathcal{A}
\end{equation}
Hence, for each training pair ($Z, \tilde{Z}_n$), indexed by $i$, we define the contrastive loss function as:
\begin{equation}
    \mathcal{L}_{\textrm{contrast}}^{(i)} = \frac{1}{T} \sum\limits_{t=1}^T -\log \Bigg(\frac{e^{(z_t \cdot \tilde{z}_{n_{\tilde{t}}}/\tau)}}{e^{(z_t \cdot \tilde{z}_{n_{\tilde{t}}}/\tau)}+ \sum\limits_{k=1}^{K} e^{(z_t \cdot z_{k}/\tau)}}\Bigg),
\label{contra_loss}
\end{equation}
where $z_k$ are $K$ negative embeddings randomly chosen from other training pairs in a batch, i.e., embeddings from terms $w' \neq w$.

\textbf{Tokenization.} We further discretize $Z$ and $\tilde{Z}_n$ using a Vector Quantizer (VQ), denoted as $q(\cdot)$, to obtain their corresponding discrete sequences $\Hat{Z}$ and $\Hat{Z}_n$, respectively. The VQ utilizes a trainable codebook $C$ consisting of $K$ $d$-dimensional discrete codewords, denoted as $C = \{c_1, c_2, \dots, c_K\}$. Specifically, the function $q: Z \rightarrow \Hat{Z}$ converts a sequence of continuous encoder representations $Z=\{z_t\}_{t=1}^T$ into $\Hat{Z} = \{\Hat{z}_t\}_{t=1}^T$ by mapping each $z_t$ to its nearest codeword $c_k \in C$ in terms of cosine similarity:
\begin{equation}
    \Hat{z}_t = \frac{c_{k^*}}{\left \Vert c_{k^*} \right \Vert_2} , \textrm{where}\, k^*= \arg \max_{c_k \in C} \bigg (z_t\cdot \frac{c_k}{ \left \Vert c_k \right \Vert_2}\bigg ), 
\end{equation}

We also add a commitment loss to ensure the embeddings are closely aligned with their corresponding discrete representations, and we define the loss for the $i^{th}$ pair as:

%+  z_{n_t} \cdot \Hat{z}_{n_t}
\begin{equation}
    \mathcal{L}_{\textrm{commit}}^{(i)} = -\frac{1}{T}\sum\limits_{t=1}^T z_t \cdot \Hat{z}_t 
\end{equation}
\textbf{Robust discrete tokens.} To ensure robustness in tokenization, we aim to map frame-level embeddings from any given anchor-positive pair ($z_t$ and $\tilde{z}_{n_{\tilde{t}}}$) to the same codeword consistently. To achieve this, we define a robust consistency loss for the $i^{th}$ pair as follows: 
\begin{equation}
    \mathcal{L}_{\textrm{robust}}^{(i)} = \frac{1}{|\mathcal{A}|} \sum_{(t, \tilde{t}) \in \mathcal{A} }\mathcal{L}(z_t, \tilde{z}_{n_{\tilde{t}}}) + \mathcal{L}(\tilde{z}_{n_{\tilde{t}}}, z_t)\, 
\end{equation}
where $\mathcal{L}(z_t, \tilde{z}_{n_{\tilde{t}}})$ denotes the cross-entropy loss between two distributions associated with $z_t$ and $\tilde{z}_{n_{\tilde{t}}}$. This loss is defined as: 

\begin{equation}
    \mathcal{L}(z_t, \tilde{z}_{n_{\tilde{t}}}) = -\sum_{k=1}^K p(z_t|c_k)\log\bigg(\frac{\exp ((\tilde{z}_{n_{\tilde{t}}}\cdot c_k)/\tau')}{\sum_{k'}\exp ((\tilde{z}_{n_{\tilde{t}}}\cdot c_{k'})/\tau')}\bigg),
    \label{eq 5}
\end{equation}
where $p(z_t|c_k)$ represents the probability of $z_t$ being assigned to codeword $c_k$, and $\tau'$ is a parameter that controls the sharpness of the distribution. We compute $\mathcal{L}(\tilde{z}_{n_{\tilde{t}}}, z_t)$ vice versa. \\

\noindent\textbf{Balanced codebook.} The trainable codebooks are observed to suffer from the index collapse problem, i.e. $p(z|c_k)$ in Eq. \ref{eq 5} becomes skewed distribution. To address this, we propose a novel balanced clustering objective to ensure that the frame-level embeddings $z$ are uniformly distributed across all codewords $c_k$ in each training batch. The objective is formulated as follows:
\begin{equation}
    \begin{aligned}
        & \max_p \mathbb{E}_z \bigg[\sum_{k=1}^{K}p(z|c_k) s_k(z) \bigg], \text{where}\,\ s_k(z) = z \cdot \frac{c_k}{\left \Vert c_k \right \Vert_2} \\
        & \text{subject to} \quad \mathbb{E}_z[p(z|c_k)] = \frac{1}{K} \forall k    \\
    \end{aligned}
\label{eq:placeholder_label}
\end{equation}
% The above formulation can be interpreted as an instance of an optimal transport (OT) problem, which can be efficiently solved in a nearly linear time using the Sinkhorn-Knopp algorithm \cite{cuturi2013sinkhorn}. We compute the cost of mapping the $t^{th}$ sample $z_t$ to $c_k$ as -$s_k(z_t)$ in the OT formulation. 
The above formulation can be interpreted as a case of the optimal transport (OT) problem, which can be effectively computed in almost linear time with the Sinkhorn-Knopp algorithm \cite{cuturi2013sinkhorn}. In the OT framework, we define the cost of assigning the $t^{th}$ sample $z_t$ to $c_k$ as -$s_k(z_t)$. The resulting solution provides $p(z_t|c_k)$, which we use in Eq. \ref{eq 5}. \\

\noindent \textbf{Training objective.} Finally, we train our model using the total loss $\mathcal{L}$ computed over training batch of size $B$ as:
\begin{equation}
    \mathcal{L} = \frac{1}{B}\sum_{i=1}^{B} \mathcal{L}_{\textrm{contrast}}^{(i)} + \lambda_1\mathcal{L}_{\textrm{robust}}^{(i)} + \lambda_2\mathcal{L}_{\textrm{commit}}^{(i)},
\end{equation}
where $\lambda_1$ and $\lambda_2$ controls tradeoff between loss components. 

\subsection{Indexing and Retrieval}

We adopt the fast progressive search strategy proposed in our previous work \cite{singh25d_interspeech}. Given a set $\mathcal{D}$ of audio tracks, each track $a_i \in \mathcal{D}$ is divided into overlapping segments of length $l$\,s  with a hop size of $h$\,s. For each segment, we compute its representation $Z=\{z_t\}_{t=1}^{T}$, followed by its corresponding tokenized representation $ Z' = \{i_t\}_{t=1}^T$, where each $i_t$ denotes the index of the nearest codeword for $z_t$ in the codebook $C$, with  $i_t \in \{1, 2, \dots, K\}$. Subsequently, we construct a TF-IDF representation for each tokenized sequence \( Z' \) in the database and index them using IVF-PQ \cite{jegou2010product} for fast retrieval.
% To further enhance retrieval efficiency and precision, we employ a multistage search strategy, described in Appendix, which progressively filters candidates at each stage to refine the results.

To further enhance retrieval efficiency and precision, we perform retrieval in multiple stages, progressively filtering candidates at each stage to refine the results and improve precision. Given a query $q$, we first compute its token representation $Z'_q$ and its corresponding TF-IDF representation. In the initial stage, we retrieve a set of candidate matches $\mathcal{P}_1$ using the index. In the second stage, we further refine $\mathcal{P}_1$ by computing the Jaccard similarity between $Z'_q$ and $Z' \in \mathcal{P}_1$, resulting in a filtered set $\mathcal{P}_2$. Lastly, we apply an edit distance-based filtering on $\mathcal{P}_2$ to obtain the final set of best-matches $\mathcal{P}_3$. The edit distance is used to incorporate temporal information, which is absent in the Jaccard similarity computation performed in the previous stage. Our retrieval process is limited to this stage for efficiency reasons. However, to further improve precision, DTW can be applied as additional filters using the discrete representation $Z'$ and the continuous representation $Z$ in succession.
\label{sec:method}

\section{Experimental Setup}

\subsection{Databases}
We trained the model on the LibriSpeech \textit{train-clean-360} subset and used the \textit{test-clean} subset for validation. To construct the speech archive and queries, we selected the \textit{train-clean-100} subset, which contains approximately 100 hours of spoken content. This setup ensures that the evaluation is always conducted on speakers unseen during training our model. To assess cross-dataset generalization, we also evaluated our model on the TIMIT dataset \cite{garofolo1993darpa}. We employed Montreal Forced Aligner \cite{mcauliffe2017montreal} to obtain word alignments for both datasets. To emulate realistic acoustic conditions, the training data were augmented with a diverse range of background noise types and reverberation effects by incorporating noise samples and room impulse responses (RIRs) from the MUSAN corpus \cite{snyder2015musan}. For evaluation, we employed RIRs from the Aachen Impulse Response Database \cite{jeub2009binaural} and noise clips from the ETSI background noise database \cite{transmission1994speech}, both of which contain recordings of real-world acoustic environments such as offices, public spaces, and street settings.

\subsection{Evaluation}
We created two distinct query sets, each containing 100 unique terms
\begin{itemize} %[leftmargin=1pt]
    \item In-Vocabulary – IV: This set includes terms whose text forms exist in the training data, but the query utterances are spoken by unseen speakers.
    \item Out-of-Vocabulary – OOV: This set contains terms whose text forms and speakers are absent from the training data.
\end{itemize}
We evaluated the  spoken term detection baselines using the Mean Term Weighted Value (MTWV) \cite{shah2020cross} as the primary performance metric. Also, to assess the robustness of speech tokenizers against speaker variations and audio distortions, we measured the Jaccard similarity \cite{liu2009learning} between the tokenized representations of different utterances of the same spoken term.
% Also, we statistically computed thresholds for retrieval for different baselines.

\subsection{Baselines}
We evaluated ASR-based baselines for STD tasks, including HuBERT \cite{hsu2021hubert} and WavLM \cite{chen2022wavlm}, by extracting posterior tokens from these models. Also, we evaluated speech tokenizers that generate semantic speech tokens, such as WavLM, SpeechTokenizer \cite{zhang2023speechtokenizer}, and BEST-STD \cite{singh2025best}, for STD tasks. For WavLM, we used the pretrained KMeans model available in the SpeechBrain toolkit \cite{ravanelli2021speechbrain}, which was trained on features extracted from the 23rd layer of the large variant of WavLM. For SpeechTokenizer \cite{zhang2023speechtokenizer}, we extracted the semantic tokens generated by the first quantizer in its residual vector quantization (RVQ) stack.

\subsection{Implementation details}
The input audio segments were set to 1\,s ($l$), covering approximately 93\% of spoken terms in the database to ensure sufficient term coverage during training. We extracted 16 MFFCs along with their first and second derivatives, computed over 25ms windows with a 10ms frameshift. The encoder consisted of 8 bidirectional Mamba layers, followed by a projection layer mapping the output to a 128-dimensional embedding space, totaling 8.1M trainable parameters. Hyperparameters $\tau$ and $\tau'$ were fixed at 0.1, while $\lambda_1$ and $\lambda_2$ were set to 1 and 10, respectively. The model was trained for 740k steps using the Adam optimizer with a learning rate of $5 \times 10^{-4}$ and a batch size of 96. During training, clean speech was mixed with noise at SNRs uniformly sampled from 0–10 dB. Evaluation was performed at fixed SNRs in range -5 to 20dB, including values outside the training range to assess generalisation to unseen noise levels.

\begin{table*}[ht]
\vspace{-6pt}
\caption{The average Jaccard similarity ($\uparrow$) between the tokenized representations of utterance pairs across various distortion conditions.}
\centering
\resizebox{0.5\textwidth}{!}{
\begin{tabular}{|c|c|c|ccccc|ccccc|}
\hline
\textbf{Model}                                 & \textbf{Tokens} & \textbf{Clean} & \multicolumn{5}{c|}{\textbf{Noise}}                                           & \multicolumn{5}{c|}{\textbf{Noise+Reverb ($t_{60}=0.7s$)}}                             \\ \hline
                                               &                 &                & \textbf{-5dB} & \textbf{0dB}  & \textbf{5dB}  & \textbf{10dB} & \textbf{15dB} & \textbf{-5dB} & \textbf{0dB}  & \textbf{5dB}  & \textbf{10dB} & \textbf{15dB} \\ \hline
\multicolumn{1}{|l|}{\textbf{ASR Posteriors:}} &                 &                &               &               &               &               &               &               &               &               &               &               \\
HuBERT-Large \cite{hsu2021hubert}                                   & 32              & 0.73           & 0.46          & 0.59          & 0.67          & 0.71          & 0.72          & 0.24          & 0.37          & 0.49          & 0.58          & 0.64          \\
WavLM-Large \cite{chen2022wavlm}                                   & 32              & 0.72           & 0.62          & 0.67          & 0.70          & 0.71          & 0.71          & 0.52          & 0.60          & 0.65          & 0.68          & 0.70          \\
\multicolumn{1}{|l|}{\textbf{Speech Tokens:}}  &                 &                &               &               &               &               &               &               &               &               &               &               \\
SpeechTokenizer \cite{zhang2023speechtokenizer}                               & 1024            & 0.45           & 0.09          & 0.12          & 0.15          & 0.18          & 0.19          & 0.03          & 0.04          & 0.05          & 0.07          & 0.08          \\
WavLM-Large \cite{chen2022wavlm}                                   & 1000            & 0.40           & 0.18          & 0.19          & 0.21          & 0.22          & 0.23          & 0.16          & 0.17          & 0.18          & 0.18          & 0.20          \\

BEST-STD \cite{singh2025best}                                   & 1024            & 0.72           & 0.21          & 0.29          & 0.42          & 0.60          & 0.65          & 0.19          & 0.22          & 0.38          & 0.55          & 0.62          \\
\hline
Ours - Transformer                             & 1024            &     0.78           &   0.67            & 0.73               &  0.75             &   0.77            & 0.77             &     0.57          &    0.64           & 0.68              & 0.72       &0.73       \\ 
BEST-STD 2.0                                 & 1024            & \textbf{0.86}  & \textbf{0.72} & \textbf{0.78} & \textbf{0.81} & \textbf{0.83} & \textbf{0.84} & \textbf{0.61} & \textbf{0.69} & \textbf{0.74} & \textbf{0.77} & \textbf{0.79} \\
\hline
\end{tabular}}
\label{jacc_sim}
\end{table*}

\begin{table*}[ht]
\centering
\vspace{-6pt}
\caption{Spoken Term Detection MTWV ($\uparrow$) under various distortion conditions on LibriSpeech (left) and TIMIT (right).}
\resizebox{\textwidth}{!}{%
\begin{tabular}{|l|cccccc|cccccc||cccccc|cccccc|}
\hline
\multirow{3}{*}{\textbf{Model}} 
& \multicolumn{12}{c||}{\textbf{LibriSpeech}} 
& \multicolumn{12}{c|}{\textbf{TIMIT}} \\ \cline{2-25}
& \multicolumn{6}{c|}{\textbf{IV}} & \multicolumn{6}{c||}{\textbf{OOV}}
& \multicolumn{6}{c|}{\textbf{IV}} & \multicolumn{6}{c|}{\textbf{OOV}} \\ \cline{2-25}
& \textbf{-5dB} & \textbf{0dB} & \textbf{5dB} & \textbf{10dB} & \textbf{15dB} & \textbf{20dB}
& \textbf{-5dB} & \textbf{0dB} & \textbf{5dB} & \textbf{10dB} & \textbf{15dB} & \textbf{20dB}
& \textbf{-5dB} & \textbf{0dB} & \textbf{5dB} & \textbf{10dB} & \textbf{15dB} & \textbf{20dB}
& \textbf{-5dB} & \textbf{0dB} & \textbf{5dB} & \textbf{10dB} & \textbf{15dB} & \textbf{20dB} \\
\hline
\multicolumn{25}{|l|}{\textbf{Noise}} \\ \hline
\textbf{ASR Posteriors:} &  & & & & &  & & & & & & &  & & & & &  & & & & & & \\ 
HuBERT-Large \cite{hsu2021hubert}  & 0.13 & 0.21 & 0.30 & 0.40 & 0.47 & 0.47 & 0.16 & 0.27 & 0.34 & 0.40 & 0.41 & 0.43
& 0.14 & 0.22 & 0.31 & 0.43 & 0.49 & 0.51 & 0.16 & 0.28 & 0.37 & 0.43 & 0.44 & 0.46 \\ 
WavLM-Large \cite{chen2022wavlm}  & 0.31 & 0.36 & 0.43 & 0.52 & 0.55 & 0.58 & 0.29 & 0.37 & 0.41 & 0.42 & 0.43 & 0.45
& 0.33 & 0.35 & 0.44 & 0.52 & 0.55 & 0.61 & 0.33 & 0.41 & 0.46 & 0.47 & 0.49 & 0.50 \\
\textbf{SpeechTokens:} &  & & & & &  & & & & & & & & & & &  & & & & & & & \\ 
SpeechTokenizer \cite{zhang2023speechtokenizer}  & 0.14 & 0.27 & 0.39 & 0.49 & 0.52 & 0.53 & 0.13 & 0.21 & 0.30 & 0.42 & 0.48 & 0.49
& 0.15 & 0.28 & 0.42 & 0.53 & 0.56 & 0.57 & 0.15 & 0.26 & 0.34 & 0.43 & 0.48 & 0.52 \\
WavLM-Large \cite{chen2022wavlm} & 0.17 & 0.34 & 0.40 & 0.53 & 0.55 & 0.55 & 0.17 & 0.25 & 0.35 & 0.43 & 0.47 & 0.49
& 0.19 & 0.38 & 0.44 & 0.57 & 0.59 & 0.61 & 0.19 & 0.29 & 0.35 & 0.46 & 0.47 & 0.51 \\
BEST-STD \cite{singh2025best} & 0.27 & 0.35 & 0.43 & 0.50 & 0.57 & 0.62 & 0.22 & 0.29 & 0.37 & 0.44 & 0.49 & 0.54
& 0.29 & 0.38 & 0.47 & 0.54 & 0.62 & 0.66 & 0.25 & 0.33 & 0.40 & 0.49 & 0.50 & 0.56 \\ 
\hline
Ours-Transformer & 0.51 & 0.58 & 0.61 & 0.65 & 0.67 & 0.67 & 0.50 & 0.56 & 0.60 & 0.62 & 0.64 & 0.65
& 0.55 & 0.62 & 0.66 & 0.73 & 0.74 & 0.75 & 0.52 & 0.60 & 0.64 & 0.66 & 0.68 & 0.69 \\
\textbf{BEST-STD 2.0} & \textbf{0.58} & \textbf{0.64} & \textbf{0.72} & \textbf{0.75} & \textbf{0.77} & \textbf{0.77} & \textbf{0.51} & \textbf{0.62} & \textbf{0.65} & \textbf{0.67} & \textbf{0.68} & \textbf{0.68}
& \textbf{0.60} & \textbf{0.67} & \textbf{0.78} & \textbf{0.80} & \textbf{0.81} & \textbf{0.82} & \textbf{0.53} & \textbf{0.63} & \textbf{0.67} & \textbf{0.69} & \textbf{0.70} & \textbf{0.71} \\ \hline
\multicolumn{25}{|l|}{\textbf{Noise + Reverberation ($t_{60}=0.7s$)}} \\ \hline

\textbf{ASR Posteriors:} &  & & & & &  & & & & & & &  & & & & &  & & & & & & \\  
HuBERT-Large \cite{hsu2021hubert}  & 0.02 & 0.06 & 0.09 & 0.13 & 0.21 & 0.24 & 0.02 & 0.07 & 0.12 & 0.20 & 0.26 & 0.29
& 0.03 & 0.08 & 0.12 & 0.23 & 0.25 & 0.27 & 0.08 & 0.15 & 0.24 & 0.26 & 0.28 & 0.30 \\
WavLM-Large \cite{chen2022wavlm}  & 0.11 & 0.18 & 0.24 & 0.30 & 0.32 & 0.36 & 0.15 & 0.22 & 0.29 & 0.31 & 0.35 & 0.37
& 0.12 & 0.21 & 0.23 & 0.35 & 0.37 & 0.39 & 0.18 & 0.24 & 0.31 & 0.35 & 0.39 & 0.41 \\
\textbf{SpeechTokens:} &  & & & & &  & & & & & & & & & & &  & & & & & & & \\ 
SpeechTokenizer \cite{zhang2023speechtokenizer} & 0.03 & 0.05 & 0.11 & 0.14 & 0.18 & 0.20 & 0.02 & 0.04 & 0.06 & 0.11 & 0.13 & 0.16
& 0.05 & 0.12 & 0.18 & 0.19 & 0.23 & 0.23 & 0.07 & 0.11 & 0.14 & 0.18 & 0.21 & 0.23 \\
WavLM-Large \cite{chen2022wavlm}  & 0.06 & 0.12 & 0.19 & 0.25 & 0.34 & 0.39 & 0.04 & 0.07 & 0.14 & 0.21 & 0.27 & 0.31
& 0.08 & 0.16 & 0.23 & 0.26 & 0.30 & 0.36 & 0.10 & 0.17 & 0.23 & 0.25 & 0.29 & 0.30 \\
BEST-STD \cite{singh2025best} & 0.18 & 0.26 & 0.34 & 0.40 & 0.46 & 0.51 & 0.13 & 0.20 & 0.27 & 0.33 & 0.39 & 0.43
& 0.20 & 0.28 & 0.36 & 0.44 & 0.49 & 0.54 & 0.17 & 0.26 & 0.33 & 0.34 & 0.42 & 0.48 \\
\hline
Ours-Transformer & 0.41 & 0.50 & 0.55 & 0.58 & 0.58 & 0.60 & 0.40 & 0.46 & 0.52 & 0.55 & 0.57 & 0.57
& 0.43 & 0.52 & 0.56 & 0.62 & 0.63 & 0.64 & 0.41 & 0.51 & 0.55 & 0.56 & 0.58 & 0.59 \\
\textbf{BEST-STD 2.0} & \textbf{0.45} & \textbf{0.53} & \textbf{0.61} & \textbf{0.67} & \textbf{0.68} & \textbf{0.68} & \textbf{0.40} & \textbf{0.50} & \textbf{0.56} & \textbf{0.58} & \textbf{0.61} & \textbf{0.62}
& \textbf{0.47} & \textbf{0.54} & \textbf{0.63} & \textbf{0.67} & \textbf{0.70} & \textbf{0.71} & \textbf{0.43} & \textbf{0.54} & \textbf{0.60} & \textbf{0.61} & \textbf{0.65} & \textbf{0.66} \\ \hline
\end{tabular}%
}
\label{main_res}
\end{table*}

\section{Results}
\subsection{Analysis of Speech Tokens}
We assess token consistency using 5k cross-speaker spoken-term pairs and compute their Jaccard similarity across all acoustic conditions. Table \ref{jacc_sim} shows that our tokenizer produces the most consistent tokens across all conditions, outperforming every baseline, including K-Means tokens, ASR posterior tokens from WavLM, and the earlier BEST-STD system. Even under unseen low-SNR ($\leq$ 5 dB) and highly reverberant conditions, our tokens maintain high Jaccard similarity, whereas all baselines degrade sharply. Notably, in clean conditions, our tokens still improve similarity over BEST-STD by 10\%, indicating that the gains are not limited to noise robustness but also reflect stronger speaker invariance and more efficient tokenization. These gains stem from our noise-augmented training and optimal-transport objective, which together mitigate codebook collapse and yield well-separated, discriminative token representations. This analysis provide direct evidence that our tokens retain speaker-invariant behavior in addition to noise robustness.

We further compare tokens extracted from our Transformer-based encoder, trained within the same framework, to those from the BiMamba encoder. Despite both models having similar sizes, the Transformer-based tokens exhibit lower efficiency. We attribute this performance gap to the rotary positional encoding used in the Transformer, which may capture unnecessary temporal variations, whereas the BiMamba model provides more effective temporal modeling.

\subsection{Codebook analysis}
We further evaluate the effectiveness of our proposed balanced clustering objective (Eq. 8) in ensuring uniform token utilization. To assess token balance, we compute the normalized entropy of the codebook $C$, defined as:
\begin{equation}
    \mathcal{H}(C) = -\frac{1}{\log K}\sum_{k=1}^{K}p_k\log(p_k)
\end{equation}
As shown in Table \ref{tokens_balance}, our approach consistently achieves a normalized entropy close to 1, indicating near-perfect balance across different codebook sizes, ranging from 1024 to 4096.  In contrast, KMeans-based tokenization \cite{singh2025best} yields substantially lower entropy that drops further as the codebook size increases. Furthermore, we encountered codebook collapse when employing a trainable codebook regularized with KL divergence \cite{baevski2020wav2vec}, despite extensive optimization efforts. These results demonstrate that our approach effectively overcomes the codebook collapse problem, enabling stable and scalable learning of large speech-token codebooks.

\begin{table}[h]
\vspace{-6pt}
\caption{Normalized entropy of the codebook for different codebook sizes.}
\centering
\resizebox{0.8\columnwidth}{!}{
\begin{tabular}{|c|c|ccc|}
\hline
\textbf{Models} & \textbf{Tokens type} & \textbf{1024} & \textbf{2048} & \textbf{4096} \\ \hline
KL Divergence \cite{baevski2020wav2vec}       & Learnable         & 0.63          & 0.50          & 0.38          \\
BEST-STD \cite{singh2025best}       & Kmeans         & 0.76          & 0.65          & 0.43          \\
BEST-STD 2.0  & Learnable            & \textbf{0.98}          & \textbf{0.97}          & \textbf{0.96}          \\ \hline
\end{tabular}}
\label{tokens_balance}
\end{table}

\subsection{Retrieval}

Table \ref{main_res} presents the retrieval performance of various methods across different noisy environments. Our approach consistently outperforms every baseline even under severe noise and reverberation, demonstrating strong robustness. ASR-based models exhibit significant retrieval failures due to transcription errors, particularly when limited context leads to misrecognitions. Notably, our method surpasses WavLM-based approaches, including both KMeans-derived tokens and ASR posteriors, despite WavLM being explicitly trained on large-scale datasets for noise-robustness. Similar trends are observed on the TIMIT database across different acoustic conditions. 
% This efficiency makes it particularly advantageous for practical deployment, enabling faster and more cost-effective speech retrieval. 

While our Transformer-based model, trained within the same framework, achieves competitive results, it falls short of the Bidirectional Mamba model in noisy and reverberant settings. This performance gap can be attributed to the Transformer's less effective temporal modeling compared to Bidirectional Mamba, which processes temporal dependencies in a linear fashion, making it more suited for our context. BEST-STD also underperforms due to tokenization inefficiencies, where the lack of token diversity leads to a higher false-positive rate. In contrast, our method ensures well-separated and distinctive tokens, resulting in improved retrieval accuracy. Importantly, our approach demonstrates strong performance on OOV terms, highlighting the compositional nature of the generated tokens. This suggests that our method can generalize effectively beyond seen vocabulary, reinforcing its applicability in real-world spoken term detection scenarios.

We benchmark retrieval latency using an in-memory search on an Intel Xeon Platinum 8268 CPU. Our system retrieves the top-10 matches for a spoken query in $\sim$1.2s on average, compared to $\sim$3.4s for BEST-STD, which relies on an inverted index. This shows a roughly $3\times$ speedup. We attribute this gain to the proposed TF-IDF-based retrieval strategy, which reduces search overhead while maintaining retrieval accuracy. 

Due to space constraints, additional results including the ablation study and qualitative analysis are available on our project webpage
\footnote{\label{myfootnote}\url{https://github.com/anupsingh15/BEST-STD2.0}}.

\section{Conclusion}
In this paper, we present a robust speech tokenization method that performs effectively under challenging conditions such as noise and reverberation. Furthermore, by incorporating a balanced token learning strategy, our method offers a scalable solution to the codebook collapse problem, ensuring efficient and stable training even with large codebooks. Our approach underscores a paradigm shift toward token-based speech representations that enable text-like indexing and search, thereby bridging the methodological gap between spoken content retrieval and natural language processing. 

\section{Acknowledgment}
This research was supported by the Research Foundation Flanders (FWO) under grant S004923N of the SBO programme .

%\vfill
%\pagebreak

\bibliographystyle{IEEEbib}
\bibliography{strings,refs}

\end{document}